\documentclass[twocolumn]{aastex61}
\accepted{August 8, 2018}
\submitjournal{ApJL}
\shorttitle{Discovery of Two New GCs}
\shortauthors{Ryu \& Lee}
\begin{document}

\title{Discovery of Two New Globular Clusters in the Milky Way}

\correspondingauthor{Myung Gyoon Lee}
\email{mglee@astro.snu.ac.kr}

\author{Jinhyuk Ryu}
\affil{Astronomy Program, Department of Physics and Astronomy, Seoul National University, Korea}
\email{ryujh@astro.snu.ac.kr}

\author{Myung Gyoon Lee}
\affiliation{Astronomy Program, Department of Physics and Astronomy, Seoul National University, Korea}

\begin{abstract}
The spatial distribution of known globular clusters (GCs) in the Milky Way shows that the current census of GCs is incomplete in the direction of the Galactic plane. We present the discovery of two new GCs located close to the Galactic plane in the sky. These two GCs, RLGC 1 and RLGC 2, were discovered serendipitously during our new cluster survey \citep{ryu18} based on near-Infrared and mid-Infrared survey data. The two GCs show a grouping of resolved stars in their $K$ band images and a presence of faint diffuse light in their outer regions in the \textit{WISE W1} band images. They also show prominent red giant branches (RGBs) in their $K$ \replaced{--}{vs.} $(J-K)$ color-magnitude diagrams (CMDs). We determine structural parameters of the two GCs using King profile fitting on their $K$ band radial number density profiles. The determined values are consistent with those of known GCs. Finally, we determine the distances, metallicities, and reddenings of the two GCs using the isochrone fitting on their CMDs. For the fitting, we assume that the ages of the two GCs are 12.6 Gyr and the brightest RGB stars of each cluster correspond to the tip of the RGB. Distances and metallicities of the two GCs are estimated to be $d=28.8\pm4.3$ kpc and $\textrm{[Fe/H]}=-2.2\pm0.2$ for RLGC 1 and $d=15.8\pm2.4$ kpc and $\textrm{[Fe/H]}=-2.1\pm0.3$ for RLGC 2. These results show that the two GCs are located at the far-half region of the Milky Way and they may belong to the halo of the Milky Way. 
\end{abstract}


\section{Introduction}
The current census of the Milky Way Globular Clusters (GCs) is incomplete in the direction of the Galactic plane.
The severe extinction of the Galactic plane prevents us from finding GCs located in the far-half region of the Milky Way (behind the center of the Milky Way). 
According to the catalog of the Milky Way GCs in \citet[2010 edition]{har96}, there are 57 GCs at $|Z|<1$ kpc. 
While 43 of these GCs 
are in the close-half of the Galaxy, only 14 GCs are in the far-half region.

In addition to the GCs in the Harris catalog, many new GCs \deleted{and GC candidates} in the Milky Way were discovered since 2005. The total number of these objects amounts to \replaced{117}{55}  
:Willman 1 \citep{wil05}, FSR 584 \citep{bic07}, FSR 1767 \citep{bon07}, FSR 190 \citep{fro08a}, FSR 1716 \replaced{\citep{fro08b}}{\citep[=VVV--CL005;][]{fro08b, min17a}}, Pfleiderer 2 \citep{ort09}, SEGUE 3 \citep{bel10}, Mercer 5 \citep{lon11}, VVV--CL001 \citep{min11}, VVV--CL002 and CL003 \citep{mon11}, Mu\~noz 1 \citep{mun12}, Kronberger 49 \citep{ort12}, Balbinot 1 \citep{bal13}, VVV--CL110, CL128, CL131, CL143, and CL150 \citep{bor14}, Crater \citep{lae14, wei16}, Eridanus III \citep{bec15}, Kim 1 \citep{kim15a}, Kim 2 \citep{kim15b}, Laevens 3 \citep{lae15}, DES 1 \citep{luq16}, Kim 3 \citep{kim16}, Gaia 2 \citep{kop17}, Minniti 01--22 \citep{min17b}, Sagittarius II \citep{lae15, mut18}, \replaced{and Camargo 1102--1106 \citep{cam18}}{Camargo 1102 \citep{bic18, cam18}, and Camargo 1103--1106 \citep{cam18}}. 
Only $\sim30$ 
among these new objects (VVV--CL clusters, Minniti 01--22, and Camargo 1102--1106) are confirmed to be located in 
the central Galactic plane region. 
Even if we include these objects,
the numbers of the GCs in the close-half and far-half regions at $|Z|<1$ kpc are \replaced{48 and 30}{58 and 19}, respectively. 
This implies that there are more undiscovered GCs in the far-half region of the Galaxy.


Recently, \citet{ryu18} carried out a new survey of star clusters 
in the Galactic central region ($|l|<30\arcdeg$ and $|b|<6\arcdeg$) using near-Infrared (NIR) surveys and mid-Infrared (MIR) surveys, such as the Two Micron All Sky Survey (2MASS; \citealt{skr06}) and the \textit{Wide-field Infrared Survey Explorer}(\textit{WISE}; \citealt{wri10}). They found 923 new star clusters. 
During this survey, 
we serendipitously discovered two new GC candidates at $(l, b)=(336\arcdeg.87, 4\arcdeg.30)$ and $(27\arcdeg.63, -1\arcdeg.04)$: 
Ryu 059 and Ryu 879 (called RLGC 1 and RLGC 2 hereafter), which are reported in this Letter. 

This paper is organized as follows. We introduce the selection criteria for the GC candidates in Section 2. The two clusters turn out to be old GCs, thus we derive their structural parameters and distances, metallicities, and reddenings in Section 3. In Section 4, we discuss and compare spatial locations and physical parameters of the new GCs with those of known GCs. Finally, we conclude with a number estimation of undiscovered GCs in the far-half region of the Galactic plane.

\section{Discovery of RLGC 1 and 2}

\subsection{Data}
For RLGC 1, we use the 2MASS point source catalog, choosing the profile-fitted magnitudes. On the other hand, for RLGC 2, we use the UKIRT Infrared Deep Sky Survey Galactic Plane Survey (UKIDSS GPS; \citealt{law07, luc08}) data, which is much deeper than the 2MASS. Both \replaced{data}{datasets} are in VEGA magnitudes.

Unfortunately, the point source catalog of the UKIDSS GPS is significantly incomplete in the central regions of RLGC 2. Therefore, we derived point source function (PSF) magnitudes of the sources in RLGC 2 from the UKIDSS $J$ and $K$ images, using the \added{Image Reduction and Analysis Facility(IRAF)}\footnote{IRAF is distributed by NOAO, which are operated by the Association of Universities for Research in Astronomy, Inc. under contract with the National Science Foundation.}/DAOPHOT \citep{ste87}. For source detection, we used a $4\sigma$ detection threshold. We selected sources with good photometry using the sharpness distribution. These sources are matched with the point sources in the UKIDSS GPS catalog.
Using the magnitudes of the matched point sources, we transform the instrumental PSF magnitude to the standard system magnitude. The calibration errors are $\sim0.03$ mag in both $J$ and $K$ bands.

We note that $K$ band magnitudes in the UKIDSS system are known to be very similar to those in the 2MASS system \citep{hod09}: $K_{UKIDSS} = K_{2MASS} + 0.010 (J_{2MASS} - K_{2MASS})$. For the color range of $0<(J_{2MASS} - K_{2MASS})<2$, the difference between the two systems is smaller than 0.02 mag. Therefore, we do not distinguish $K$ band magnitudes between the 2MASS and UKIDSS system\added{s} in the following.

\subsection{Morphological Features}
\begin{figure*}
\epsscale{1.1}
\plotone{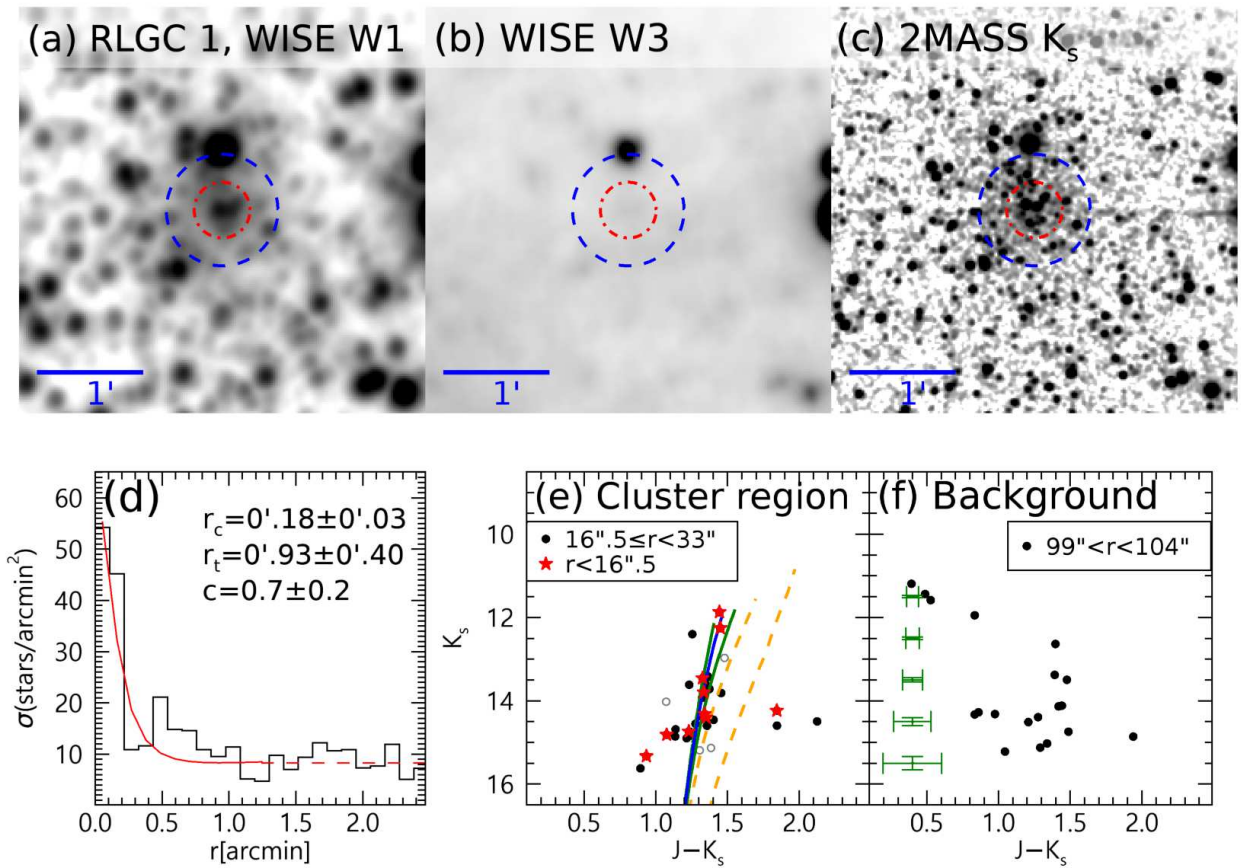}
\caption{(a) The \textit{WISE W1}, (b) \textit{W3}, and (c) 2MASS $K_s$ grayscale images of RLGC 1, respectively. The field of view is $4\arcmin\times4\arcmin$. North is up, and east is to the left. The dashed and the dotted-dashed circle represent the half-light radius ($r_h=33\arcsec$) and $0.5r_h$, respectively. (d) The radial number density profile of RLGC 1. Solid and dashed lines show the result of King profile fitting and the background level, respectively. (e) The $K_s$ vs. $(J-K_s)$ CMD of RLGC 1. Starlets and filled circles represent member stars located at $r\leq0.5r_h$ and $0.5r_h<r\leq r_h$, respectively. Gray open circles are stars removed by the statistical subtraction process. Two dashed and three solid lines are Log age $t=10.1$ (12.6 Gyr) isochrones. Their [Fe/H] values, from the left to the right, are $-2.2$, $-2.0$, $-1.8$, $-1.0$, and $+0.0$. (f) The $K_s$ vs. $(J-K_s)$ CMD of the background region. Error bars represent mean errors for given magnitudes and colors. \label{ryu059}}
\end{figure*}

\begin{figure*}
\epsscale{1.1}
\plotone{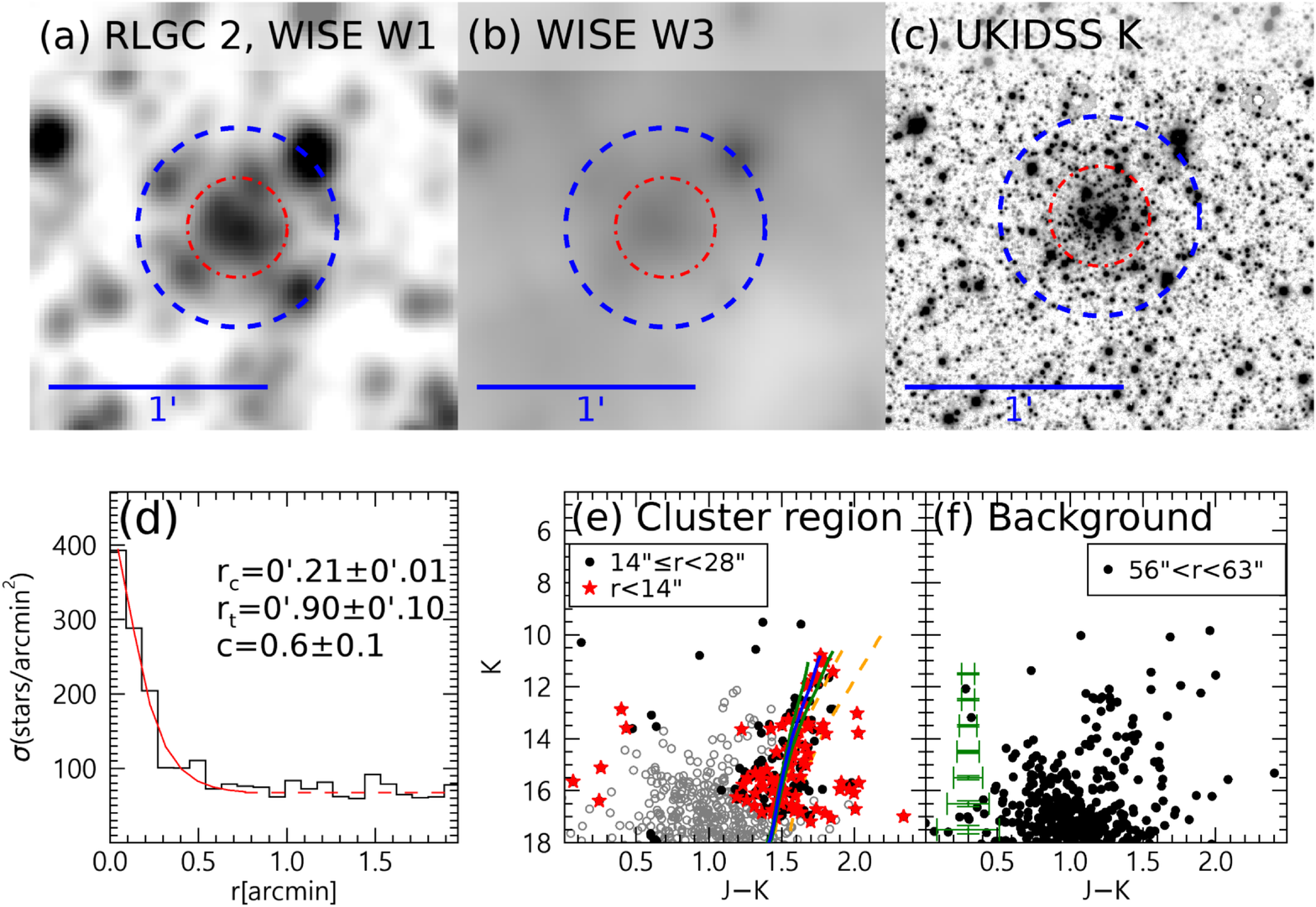}
\caption{(a) The \textit{WISE W1}, (b) \textit{W3}, and (c) UKIDSS $K$ grayscale images of RLGC 2, respectively. The field of view is $2\arcmin\times2\arcmin$. North is up, and east is to the left. Symbols and lines are the same as in Figure \ref{ryu059}. (d) The radial number density profile of RLGC 2. (e) The $K$ vs. $(J-K)$ CMD of RLGC 2. [Fe/H] values of isochrones with solid lines are $-2.4$, $-2.1$, and $-1.8$ from the left to the right, respectively. (f) The $K$ vs. $(J-K)$ CMD of the background region. \label{ryu879}}
\end{figure*}

Figure \ref{ryu059}(a), (b), and (c) show grayscale maps of the \textit{WISE W1}, \textit{W3}, and 2MASS $K_s$ band images of RLGC 1. Similar images of RLGC 2 are shown in Figure \ref{ryu879}(a), (b), and (c). The $K$ band images of these two clusters show a grouping of resolved stars. The \textit{W1} band images show a presence of faint diffuse light in the clusters. However, the \textit{W3} band images show little diffuse light features in the cluster regions, indicating that there is no dust associated with the clusters.

\subsection{Radial Number Density Profiles}
We derive the radial number density profile of the RLGC 1 region using the stars with $11\leq K_s\leq14.5$ mag, and that of the RLGC 2 region using the stars with $11\leq K\leq15$ mag. The profiles of RLGC 1 and RLGC 2 are shown in Figure \ref{ryu059}(d) and Figure \ref{ryu879}(d), respectively. 

The two clusters show strong central excesses. In the radial number density profile of RLGC 1, there is an excess at $r<0\arcmin.2$ and a slight enhancement at $0\arcmin.5<r<0\arcmin.8$. This profile becomes almost flat at $r>0\arcmin.8$. Similarly, in the case of RLGC 2, there is a central excess at $r<0\arcmin.5$. The radial number density profile of RLGC 2 becomes almost flat at $r>0\arcmin.8$. These excesses indicate that most of the bright and resolved stars in the central region\added{s} of these clusters belong to each cluster.

\subsection{Color-Magnitude Diagrams (CMDs)}

Figure \ref{ryu059}(e) and Figure \ref{ryu879}(e) display the $K$ \replaced{--}{vs.} $(J-K)$ CMDs of the resolved stars located inside the half-light radius $(r_h)$ of each cluster (which is derived in the following section). We plot the stars located at $r<0.5r_h$ by red symbols. These stars have higher probability of cluster membership compared with the stars located in the outer region. For comparison, we plot the CMDs of the background region of each cluster in Figure \ref{ryu059}(f) and Figure \ref{ryu879}(f). As background regions, we select an annular region at $99\arcsec<r<104\arcsec$ for RLGC 1 and at $56\arcsec<r<63\arcsec$ for RLGC 2. The area of each background region is the same as that of the cluster region inside the half-light radius.  

We subtract background stars from the cluster CMDs using a statistical subtraction process. This subtraction is based on the number difference of stars in the same sub-regions of the CMD between the cluster region and the background region. Details of this statistical background subtraction process is described in Appendix B of \citet{ryu18}. In Figure \ref{ryu059}(e) and Figure \ref{ryu879}(e), we plot the stars that were subtracted in this process by gray symbols. 

The CMD of RLGC 1 shows a narrow RGB feature, although the number of stars that constitute the RGB is small. The brightest part of the RGB is seen at $K_s\approx11.8$ mag and $(J-K_s)\approx1.4$. On the other hand, the CMD of RLGC 2 shows a much stronger RGB than RLGC 1. The brightest part of the RGB in this cluster is seen at $K\approx10.8$ mag and $(J-K)\approx1.8$.


In summary, based on morphological features, radial number profiles with strong central concentration, and the presence of the RGB, we conclude that RLGC 1 and RLGC 2 are old GCs.
  
 


\section{Physical Parameters of RLGC 1 and 2} 

\subsection{Structural Parameters}

We derive the structural parameters of the new GCs using King profile fitting on the background-subtracted radial number density profiles of the clusters: $\sigma=k[1/\sqrt{1+(r/r_c)^2}-1/\sqrt{1+(r_t/r_c)^2}]^2$ \citep{kin62}. In this equation, $r_c$ and $r_t$ denote the core radius and the tidal radius. The background number densities are averages of the number densities at the outer regions of each cluster, which are $72\arcsec<r<144\arcsec$ for RLGC 1 and $48\arcsec<r<96\arcsec$ for RLGC 2.
 
From this fitting, we derive the parameters for RLGC 1: $r_c=0\arcmin.18\pm0\arcmin.03$, $r_t=0\arcmin.93\pm0\arcmin.40$, and $c=0.7\pm0.2$. In the same manner, we derive structural parameters of RLGC 2: $r_c=0\arcmin.21\pm0\arcmin.01$, $r_t=0\arcmin.90\pm0\arcmin.10$, and $c=0.6\pm0.1$. The errors in these values are fitting errors.
 
We derive systematic deviations of the structural parameters using the bootstrap method with $N_{\textrm{repeat}}=3000$. The systematic parameter deviations of RLGC 1 are $\sigma_{rc}=0\arcmin.07$, $\sigma_{rt}=0\arcmin.26$, and $\sigma_{c}=0.2$ for the core radius, tidal radius, and concentration index. These systematic deviations and parameter fitting errors are similar \added{to each other}.

Similarly, we derive the deviations for RLGC 2: $\sigma_{rc}=0\arcmin.11$, $\sigma_{rt}=0\arcmin.10$, and $\sigma_{c}=0.3$. The mean value of the core radius during the bootstrap resampling is $<r_c>_{\textrm{boot}}=0.26$, which is consistent with the fitting result. However, the systematic deviation of the core radius is significantly larger than its fitting error. This means the derived core radius value is reliable, but not robust; a different sampling for the radial number density profile can make a different core radius of RLGC 2 within $\sigma_{rc}=0\arcmin.11$.

Based on the tidal radii of the new GCs, we derive the total magnitude of each cluster using circular aperture with tidal radii. Background levels are estimated using the annular region with $120\arcsec<r_{\textrm{bg}}<130\arcsec$ for both clusters.

The integrated aperture magnitude at each tidal radius is $K_{total}=7.45\pm0.02$ mag for RLGC 1 and $K_{total}=5.86\pm0.03$ mag for RLGC 2. 
Finally, we 
derive half-light radii: $r_h=32\arcsec.9\pm1\arcsec.9$ for RLGC 1 and $r_h=27\arcsec.9\pm0\arcsec.7$ for RLGC 2. These half-light radii are plotted with dashed blue line circles in the grayscale images of Figure \ref{ryu059} and Figure \ref{ryu879}.

\subsection{Distance, Metallicity and Reddening of RLGC 1 and 2} 

We determine the reddenings, distance moduli, and metallicities of the new GCs from isochrone (PARSEC; \citealt{bre12}) fitting on the CMDs. Since our photometric data of the new GCs do not reach the main-sequence turn-off points or subgiant branches, we adopt an age of 12.6 Gyr (Log (age)=10.1) for the new GCs. This age is close to the mean age of metal-poor GCs \citep[12.5 Gyr;][]{van13}. We also adopt [$\alpha\textrm{/Fe]}=+0.3$ and $\textrm{[Z/H]}=\textrm{[Fe/H]}+0.94\textrm{[}\alpha\textrm{/Fe]}$ \citep{tho03}. We use the same extinction law as adopted in \citet{bre12}: $R_V=3.1$, $A_K=0.366E(B-V)$ (2MASS), $A_K=0.353E(B-V)$ (UKIDSS), $E(J-K)=0.533E(B-V)$ (2MASS), and $E(J-K)=0.524E(B-V)$ (UKIDSS). Finally, we assume the brightest RGB star in each cluster corresponds to the tip of the RGB (TRGB).

We match visually the RGB feature of each cluster and 12.6 Gyr isochrones with varying metallicities. In particular, we tried to match the position of the TRGB and the slope of the RGB of each cluster with isochrones. The errors of the parameters are estimated by eye, considering the uncertainties in visual fitting.

Finally, we derive the values of the reddening, distance modulus, and metallicity: $E(B-V)=1.3\pm0.1$, $E(J-K)=0.7\pm0.1$, $(m-M)_0=17.3\pm0.3$, and [Fe/H]$=-2.2\pm0.2$ for RLGC 1, and $E(B-V)=1.9\pm0.2$, $E(J-K)=1.0\pm0.1$, $(m-M)_0=16.0\pm0.3$, and [Fe/H]$=-2.1\pm0.3$ for RLGC 2. The distance moduli correspond to the metric distances of $d=28.8\pm4.3$ kpc and $d=15.8\pm2.4$ kpc for RLGC 1 and RLGC 2. These results, in particular metallicities, show that these two clusters are genuine metal-poor GCs. We list all determined parameters of the new GCs in Table \ref{t1}. 

\begin{table}
\begin{center}
\caption{Fundamental Parameters of New GCs\label{t1}}
\begin{tabular}{lcc}
\tableline\tableline
 & RLGC 1 & RLGC 2\\
\tableline
$\alpha_{J2000}$ (hh:mm:ss)& 16:17:08.41 & 18:45:28.17\\
$\delta_{J2000}$ (dd:mm:ss)& --44:35:38.6 & --05:11:33.3\\
$l$ (deg)& 336.8696 & 27.6309\\
$b$ (deg)& 4.3031 & --1.0421\\
$(m-M)_0$ & $17.3\pm0.3$ & $16.0\pm0.3$\\
d (kpc) & $28.8\pm4.3$ & $15.8\pm2.4$\\
E(B-V) & $1.3\pm0.1$ & $1.9\pm0.2$\\
$\textrm{[Fe/H]}$ & $-2.2\pm0.2$ & $-2.1\pm0.3$\\
$r_c$ (arcmin) & $0.18\pm0.03$ & $0.21\pm0.01$\\
$r_c$ (pc) & $1.51\pm0.34$ & $0.97\pm0.15$\\
$r_t$ (arcmin) & $0.93\pm0.40$ & $0.90\pm0.10$\\
$r_t$ (pc) & $7.79\pm3.55$ & $4.14\pm0.78$\\
c & $0.7\pm0.2$ & $0.6\pm0.1$\\
$r_h$ (arcmin) & $0.55\pm0.03$ & $0.47\pm0.01$\\
$r_h$ (pc) & $4.59\pm0.74$ & $2.14\pm0.33$\\
$K_{total}$ (mag) & $-10.35\pm0.30$ & $-10.84\pm0.31$\\
$K_{50\arcsec}$ (mag) & $-10.28\pm0.30$ & $-10.10\pm0.31$\\
$V_{50\arcsec}$ \tablenotemark{a} (mag) & $-8.25\pm0.31$ & $-8.03\pm0.33$\\
\tableline
\end{tabular}
\tablenotetext{a}{\footnotesize$V_{50\arcsec}$ magnitude is converted from $K_{50\arcsec}$ using an equation $V-K_s =2.93+0.409$[Fe/H] \citep{coh07}.}
\end{center}
\end{table}

\section{Discussion and conclusion}

\subsection{Spatial Locations}

\begin{figure*}
\epsscale{0.8}
\plotone{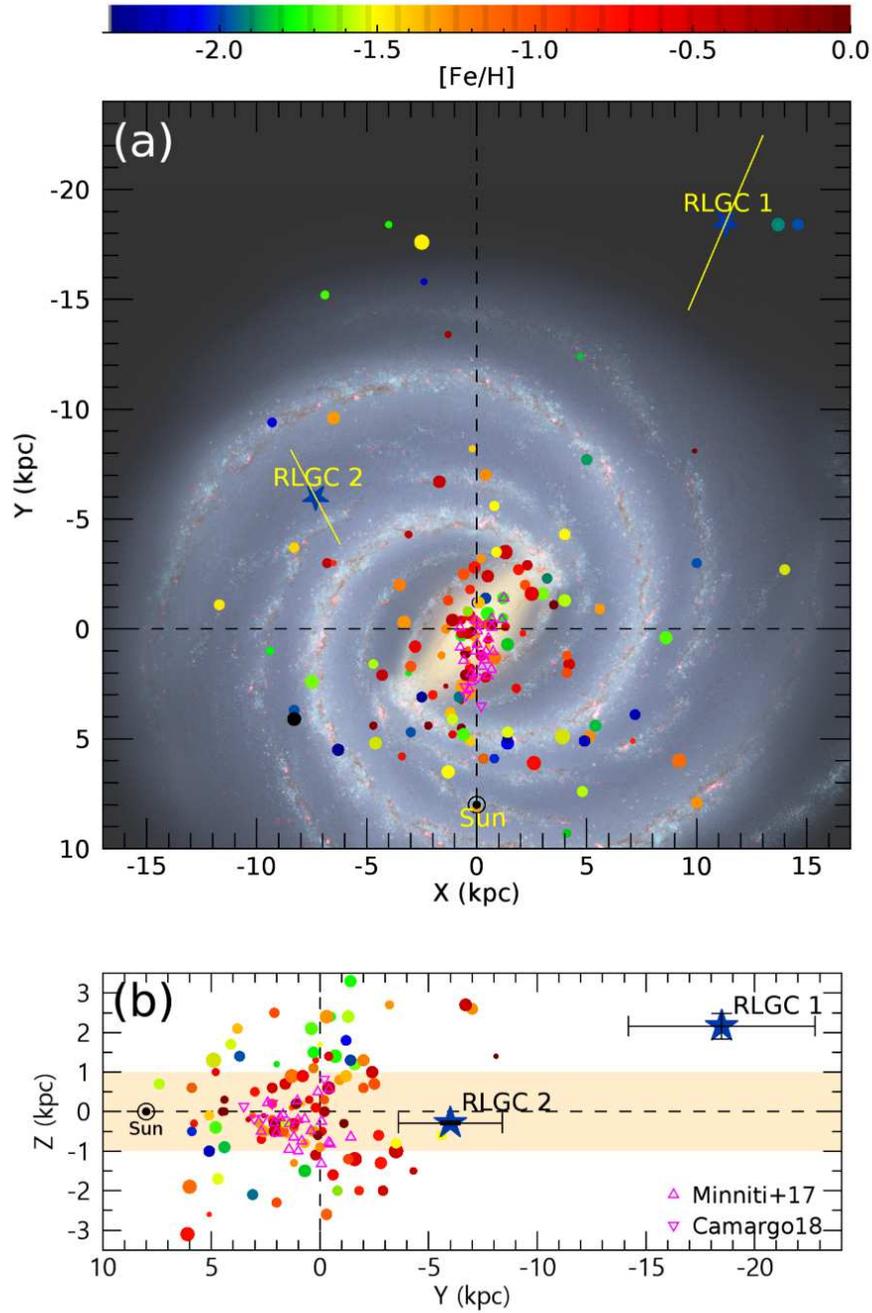}
\caption{(a) Spatial distribution of RLGC 1, RLGC 2, and the known GCs on the face-on view of the Milky Way Galaxy \citep{chu09}. Dashed lines are guide lines for the location of the Galactic center. The Sun is represented as the Sun symbol. Metallicities of GCs are color-coded as shown in the color bar at the top. The sizes of the symbol represent relative magnitudes of GCs. RLGC 1 and RLGC 2 are emphasized by the star symbol, irrespective of their magnitudes. Solid lines represent distance errors of the clusters. Open triangles represent recently reported GCs (\citealt{min17b} and \citealt{cam18}). (b) Spatial distribution of GCs on the edge-on view of the Milky Way. Symbols are the same as those in (a). The $|Z|<1$ kpc region is represented as the yellow shaded region. \label{spd}}
\end{figure*}

In Figure \ref{spd} we plot the spatial location of the new GCs in comparison with other known GCs listed in \citet{har96}, \replaced{\citet{cam18}, and Piatti (2018)}{\citet{min17b}, and \citet{cam18}}, in both the face-on view (Figure \ref{spd}(a)) and the edge-on view (Figure \ref{spd}(b)).  \added{We adopted the distances to Minniti 01--22 given by \citet{min17b} (see \citet{pia18} for other distance estimation)}. 
The two GCs are located at the far-half region of the Milky Way. Their distances from the closest neighbor GCs are 10.2 kpc (NGC 5824) and 4.5 kpc (Pal 11) for RLGC 1 and RLGC 2; practically, no neighbor GCs are found in the vicinity the new GCs.

Based on the distances and Galactic latitudes of the new GCs, we derive their vertical positions from the Galactic plane: $Z=2.2\pm0.3$ kpc for RLGC 1 and $Z=-290\pm40$ pc for RLGC 2. 
RLGC 1 that has a low metallicity is likely to be a halo GC, located above the thick disk \citep[$h_z=0.9\pm0.1$ kpc]{li17}. 
RLGC 2 is located in the thick disk. However, its low metallicity ([Fe/H]$=-2.1\pm0.3$) indicates that it must be a halo GC. Therefore, RLGC 2 may be a halo GC passing through the disk now.




\subsection{Absolute Magnitudes}
\citet{coh07} provided the integrated $K_s$ magnitude of the known GCs derived with $50\arcsec$ radius apertures. For comparison, we derive $50\arcsec$-integrated $K$ magnitudes of the two GCs. The $50\arcsec$-integrated magnitudes are $K_{50\arcsec}=-10.28\pm0.30$ mag (RLGC 1) and $K_{50\arcsec}=-10.10\pm0.31$ mag (RLGC 2), while the peak absolute magnitude of the GC luminosity function noted in \citet{coh07} is $M_K =-9.7$ mag. The magnitudes of the new GCs are 0.4--0.6 mag brighter than the peak magnitude of the known GCs in the Milky Way. Using the relation for the metal-poor GCs in \citet{coh07}: $V-K_s=2.93+0.409$[Fe/H], we estimate the $50\arcsec$-integrated $V$ magnitudes of the new GCs. The estimated magnitudes are $V_{50\arcsec}=-8.25\pm0.31$ mag (RLGC 1) and $V_{50\arcsec}=-8.03\pm0.33$ mag (RLGC 2), while the peak absolute magnitude 
is $M_V=-7.66\pm0.09$ mag \citep{dic06}. \deleted{The integrated $V$ magnitudes are also 0.4--0.6 mag brighter than the peak magnitude of the known GCs.}

\subsection{Structural Parameters}
The core radii of the two GCs are $r_c=1.51\pm0.34$ pc and $r_c=0.97\pm0.15$ pc for RLGC 1 and RLGC 2. These values are consistent with the median core radius of the known GCs, which is $median(r_c)=1.04$ pc. The half-light radii of the two GCs are also comparable with the known GCs: $r_h=4.59\pm0.74$ pc for RLGC 1, $r_h=2.14\pm0.33$ pc for RLGC 2, and $median(r_h)=3.03$ pc.

The concentration indices of the new GCs are relatively lower than those of the known GCs: $c=0.7\pm0.2$ for RLGC 1 and $c=0.6\pm0.1$ for RLGC 2 ($cf.\; median(c)=1.50$). The core radii of the new GCs are consistent with those of known GCs, hence low concentration indices would be related to tidal radii. Actually, the tidal radii of the two GCs ($r_t=7.79\pm3.55$ pc for RLGC 1 and $r_t=4.14\pm0.78$ pc for RLGC 2) are much smaller than the median tidal radius of the known GCs, $median(r_t)=28.86$ pc. This implies that the derived tidal radii of the new GCs might have been underestimated. However, complex backgrounds in the data we used prevent us from recognizing weak enhancements of number densities in outer cluster regions.

\subsection{Concluding Remarks}
Based on their morphologies, radial number density profiles, CMDs, and other determined parameters, RLGC 1 and RLGC 2 are likely to be genuine metal-poor halo GCs. However, our photometric parameters are based on uncertain assumptions: The magnitude of the brightest RGB star and the selective extinction value $R_V$. These assumptions are likely to be the origin of additional distance uncertainties. Deeper NIR photometry reaching the main-sequence turn-off point is needed to measure more accurate distances, ages, and metallicities of the two new GCs. \added{
Considering the low metallicity of these clusters, we expect that blue horizontal branch stars might be detected in deep optical CMDs reaching fainter than $V\sim23$ mag.}

Including these two new GCs, the current census of the GCs \deleted{and candidates} in the Milky Way is \replaced{$N_{\textrm{GC}}=276$}{$N_{\textrm{GC}}=214$}. The current GC numbers at $|Z|<1$ kpc are \replaced{48 and 31}{58 and 20} in the close-half region and the far-half region, respectively. Therefore, we expect that there are about \replaced{20}{30} 
undiscovered GCs in the far-half and $|Z|<1$ kpc region of the Galactic disk.

\acknowledgments
We thank Brian S. Cho for his helpful comments in improving the English in this manuscript. This work was supported by the National Research Foundation of Korea (NRF) grant funded by the Korea Government (MSIP) (No. 2017R1A2B4004632). This research has made use of the NASA/IPAC Infrared Science Archive, which is operated by the Jet Propulsion Laboratory, California Institute of Technology, under contract with the National Aeronautics and Space Administration.

\end{document}